\documentclass{article}
\usepackage{ijcai25}

\usepackage{times}
\usepackage{soul}
\usepackage{url}
\usepackage[hidelinks]{hyperref}
\usepackage[utf8]{inputenc}
\usepackage[small]{caption}
\usepackage{graphicx}
\usepackage{amsmath}
\usepackage{amsthm}
\usepackage{booktabs}
\usepackage{algorithm}
\usepackage{algorithmic}
\usepackage[switch]{lineno}

\usepackage{subfig}
\usepackage{amssymb}
\usepackage{multirow}
\usepackage{makecell}
\usepackage{setspace}
\usepackage[table]{xcolor}

\title{RoMA: Robust Malware Attribution via Byte-level Adversarial Training with Global Perturbations and Adversarial Consistency Regularization}

\author{
Yuxia Sun$^1$
\and
Huihong Chen$^1$\and
Jingcai Guo$^{2}$\and
Aoxiang Sun$^3$\and
Zhetao Li$^1$\and
Haolin Liu$^4$\\
\affiliations
$^1$College of Information Science and Technology, Jinan University\\
$^2$Department of Computing,
The Hong Kong Polytechnic University\\
$^3$College of Information and Computational Science, Jilin University\\
$^4$School of Computer Science, Xiangtan University\\
\emails
tyxsun@email.jnu.edu.cn,
chhjv@stu.jnu.edu.cn,
jc-jingcai.guo@polyu.edu.hk,
2835604014@qq.com,
liztchina@hotmail.com (corresponding author),
liu.haolin@foxmail.com
}

\begin{document}

\maketitle

\begin{abstract}
Attributing APT (Advanced Persistent Threat) malware to their respective groups is crucial for threat intelligence and cybersecurity. However, APT adversaries often conceal their identities, rendering attribution inherently adversarial. Existing machine learning-based attribution models, while effective, remain highly vulnerable to adversarial attacks. For example, the state-of-the-art byte-level model MalConv sees its accuracy drop from over 90\% to below 2\% under PGD (Projected Gradient Descent) attacks. Existing gradient-based adversarial training techniques for malware detection or image processing were applied to malware attribution in this study, revealing that both robustness and training efficiency require significant improvement. To address this, we propose RoMA, a novel single-step adversarial training approach that integrates global perturbations to generate enhanced adversarial samples and employs adversarial consistency regularization to improve representation quality and resilience. A novel APT malware dataset named AMG18, with diverse samples and realistic class imbalances, is introduced for evaluation. Extensive experiments show that RoMA significantly outperforms seven competing methods in both adversarial robustness (e.g., achieving over 80\% Robust Accuracy—more than twice that of the next-best method under PGD attacks) and training efficiency (e.g., more than twice as fast as the second-best method in terms of accuracy), while maintaining superior Standard Accuracy in non-adversarial scenarios.
The RoMA-trained model and dataset samples are publicly available at \url{https://anonymous.4open.science/r/RoMA-D767}.
\end{abstract}


\section{Introduction}

\label{sec:introduction}
Advanced Persistent Threat (APT) refers to sophisticated, covert cyberattacks typically conducted by organized groups with ties to nation-states or criminal organizations \cite{ref63}. Tracing APT groups is essential for identifying threat actors and preventing future attacks. APT malware attribution, the process of linking malware to its respective APT group, plays a critical role in this defense strategy. However, APT attackers often conceal their identities, sometimes impersonating other groups to mislead investigators, rendering APT malware attribution a challenging and adversarial task \cite{ref61,ref64}. Given the significant threat posed by Windows malware to both individuals and organizations \cite{ref5,ref6}, this study focuses on Portable Executable (PE) APT malware.


In recent years, automatic APT malware group attribution has been widely explored using dynamic and static methods. Dynamic approaches, which analyze runtime behaviors typically within virtual environments, are time-consuming and ineffective against virtual-machine escape techniques \cite{ref9,ref8}. Static methods address these issues by directly analyzing malware files, extracting features like statistical properties and program graphs from binary or disassembly code, and using classifiers like Random Forest, LSTM, and PerceiverIO \cite{ref15,ref17,ref61}. Raw-byte-based approaches further simplify the process by bypassing feature extraction and disassembly, directly processing raw byte sequences through end-to-end CNN models to capture finer details \cite{ref58}.
The state-of-the-art (SOTA) raw-byte classifier, MalConv with Global Channel Gating (GCG), uses a 1D-CNN to process raw byte sequences, achieving high accuracy in malware classification \cite{ref24}. 

\vspace{0.15cm}
\noindent\textbf{Limitations. }
Current APT malware attribution models remain vulnerable to adversarial attacks \cite{ref28}, as demonstrated by MalConv’s accuracy dropping from over 90\% to below 2\% under PGD attacks (Section \ref{sec5}). 
To enhance the robustness of raw-byte malware detectors, gradient-based adversarial training approaches have been explored. FGSM (Fast Gradient Sign Method)-based methods, including Slack-FGSM and Append-FGSM, apply adversarial perturbations to regions like slack space and file ends in PE files \cite{ref31,ref29,ref27}. While fast due to single-step training, these methods are susceptible to multi-step attacks. Recently, Lucas et al. applied multi-step PGD (Projected Gradient Descent)  training method to malware detectors, using existing perturbation techniques to mitigate high-effort PGD attacks \cite{ref55,ref56}. However, this method incurs significant training time and lack novel adversarial training techniques. Our experiments adapt these adversarial training approaches, originally developed for malware detection or image processing, to the malware attribution task, showing that both robustness and training efficiency still require substantial improvement (See Section \ref{sec5}).

\vspace{0.15cm}
\noindent\textbf{Contributions. } To address these challenges, we propose RoMA, a novel single-step adversarial training approach that efficiently trains a robust raw-byte attribution model for APT malware, offering resilience against advanced adversarial attacks, including both multi-step and optimization-based attacks. RoMA enhances FGSM-based adversarial training through two key strategies: (1) Global Perturbation (GP) strategy, which generates stronger adversarial malware by adaptively applying learned perturbation patterns (GPs) to four types of perturbation positions within malware; (2) Adversarial Consistency Regularization strategy, which introduces two loss functions, Adversarial Contrastive Loss and Adversarial Distribution Loss, to optimize the representation space by incorporating both clean and adversarial malware.

We further introduce AMG18, a novel APT malware dataset featuring varied samples and class imbalances for evaluation. Extensive experiments demonstrate that RoMA outperforms seven competitor methods, including the multi-step PGD-based training method, in both robustness and training efficiency. Under adversarial settings, such as PGD attacks, RoMA achieves over 80\% accuracy—more than twice that of the next-best method—and trains more than twice as fast. Furthermore, it achieves improved accuracy in non-adversarial scenarios.

To the best of our knowledge, this work presents the first comprehensive study on developing robust APT malware attribution models. The main contributions are as follows:
\begin{itemize}
\item Proposing RoMA, a novel single-step adversarial training approach for efficiently training a robust APT malware attribution model against advanced adversarial attacks, leveraging the Global Perturbation (GP) strategy and Adversarial Consistency Regularization.
\item Demonstrating the superiority of the RoMA-trained model in terms of adversarial robustness, clean accuracy, and training efficiency, outperforming seven competitor methods.
\item Introducing AMG18, a new APT malware dataset with diverse samples and realistic class imbalances for attribution research. The RoMA-trained model and dataset samples are publicly available \footnote{\url{https://anonymous.4open.science/r/RoMA-D767}}.
\end{itemize}

The paper is organized as follows: Section \ref{sec2} reviews related work. Section \ref{sec3} defines the problem and assumptions. Section \ref{sec4} presents the proposed RoMA approach. Section \ref{sec5} describes the experiments and results. Finally, Section \ref{sec6} concludes the paper with future research directions.

\section{Related Work}\label{sec2}

\textbf{Malware Group Attribution. } Automatic attribution of APT malware groups has been widely studied using dynamic and static approaches \cite{ref2}. Dynamic methods analyze run-time behaviors, such as function calls \cite{ref9} and network operations \cite{ref8}, but are often time-consuming and ineffective against evasion techniques like virtual machine escape. In contrast, static methods directly analyzing malware code, overcoming these limitations. Traditional static approaches rely on feature extraction, including statistical features \cite{ref15} and program graph features \cite{ref17}, and use classifiers like Random Forest \cite{ref15}, LSTM \cite{ref17} and PerceiverIO \cite{ref61}, though they require substantial manual effort for feature engineering. Other approaches automate feature extraction by processing disassembly opcode n-grams, requiring disassembly and n-gram segmentation, and use models like LSTM for classification \cite{ref18}. Raw-byte-based malware attribution simplifies the process further by bypassing feature extraction and directly capturing subtle details through end-to-end CNN models \cite{ref24,ref58}. The SOTA byte-level malware classifier, MalConv with GCG (Global Channel Gating), uses a 1D-CNN to process raw byte sequences, achieving high accuracy with innovative network designs \cite{ref24}. Despite their effectiveness, these methods remain vulnerable to adversarial attacks, which this study seeks to address.

\vspace{0.1cm}
\noindent\textbf{Adversarially Robust Malware Classification. } 
The adversarial nature between malware attacks and defenses makes malware classification models particularly vulnerable to adversarial attacks. For instance, byte-level classifiers like MalConv exhibit significant performance degradation under adversarial perturbations \cite{ref28}. To enhance robustness, recent studies have employed adversarial training techniques, generating adversarial examples using gradient-based methods that optimize functionality-preserving transformations, such as byte-level perturbations \cite{ref27} or assembly-level instruction modifications \cite{ref62}. FGSM (Fast Gradient Sign Method) offers a fast, single-step adversarial training approach. In the malware domain, Slack-FGSM and Append-FGSM apply adversarial bytes to slack space or the end of PE files \cite{ref31,ref29}, while extensions apply adversarial bytes to additional regions, such as the full DOS header or shifted sections \cite{ref27,ref28}. However, single-step FGSM methods are vulnerable to multi-step attacks. To mitigate this, multi-step adversarial training methods like PGD (Projected Gradient Descent) \cite{ref55} are employed in malware classification, although they incur considerable training overhead \cite{ref56}. Recently, Lucas et al. applied low-effort PGD attacks during adversarial training to improve the robustness of raw-byte malware detectors against high-effort PGD attacks, leveraging existing perturbation techniques without introducing new adversarial training approaches \cite{ref56}. Advanced FGSM methods from the image domain, such as NuAT (using a Nuclear-Norm regularizer) and FGSM-RS (with perturbation initialization), remain underexplored in malware classification. Our experiments adapting these image-based methods to malware classification reveal that both robustness and training efficiency still require substantial improvement.

\section{Problem Definition}\label{sec3}
\quad \textbf{Byte-level APT Malware Attribution Task:} The task is to classify APT malware into their respective APT groups by analyzing raw byte sequences. Let the input space be \( X \subseteq \{ 0,1,...,255 \}^* \) with an underlying distribution \( D \), and the label space be \( Y = \{ y_1, \dots, y_G \} \) with \( G \) APT groups. For each pair \( (x, y) \in D \), \( x \) represents a variable-length byte string, and \( y \) indicates its corresponding APT group. The goal is to find the optimal parameters \( \theta \) of the attribution model \( F(\theta): X \rightarrow Y \) that minimizes the loss function \( L(\theta, x, y) \):
\[
\underset{\theta}{\min} \, \mathbb{E}_{(x, y) \sim D} \left[ L(\theta, x, y) \right]
\]

\textbf{Attack Assumption:} This study aims to develop a robust malware attribution model resilient against adversarial attacks. Based on established threat modeling guidelines \cite{ref4}, we assume the following attacker characteristics: (1) The attacker seeks to misclassify APT malware into an incorrect APT group. (2) The attacker has complete knowledge of the classifier, including its architecture and parameters, constituting a white-box attack. (3) The attacker can modify raw bytes of the malware through functionality-preserving transformations but cannot alter the trained model. (4) The attacker can use a multi-step iterative approach, rather than a single-step attack, to enhance the attack’s intensity.

\begin{figure*}[!t]
\centering
\includegraphics[width=0.9\textwidth]{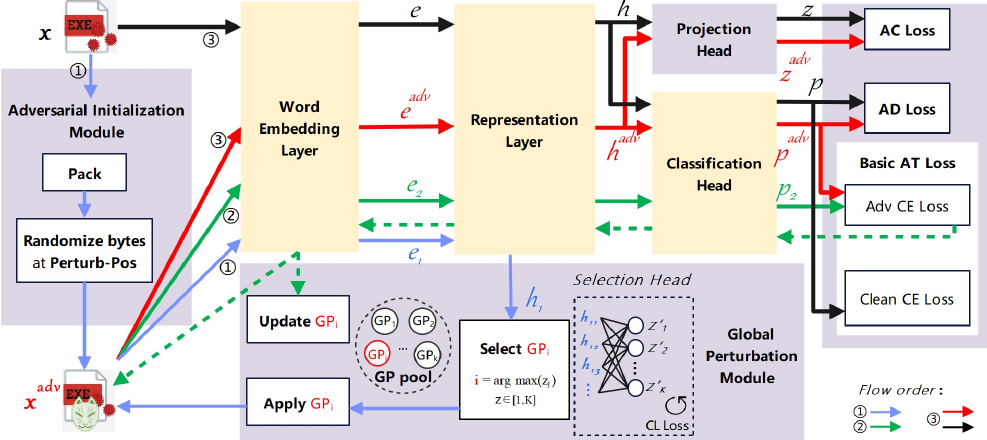}
\centering
\caption{Overview of the RoMA Training approach, with creamy-yellow components denoting the trained malware attribution model.} 
\label{fig3}
\end{figure*}

\section{Proposed Approach}\label{sec4}
\subsection{RoMA Overview}
We propose RoMA, a novel single-step adversarial training approach designed to efficiently train a robust APT malware attribution model that is resilient to multi-step attacks. \textbf{\textsl{RoMA introduces GP (Global Perturbation) to FGSM for generating enhanced adversarial malware samples and leverages adversarial consistency regularization during adversarial training to boost model robustness.}} Here, GPs are learned perturbation patterns shared by adversarial malware in the embedding space, while the regularization is applied across the prediction and projection spaces. Figure \ref{fig3} provides an overview of RoMA, illustrating its key modules and the training workflow. The creamy-yellow components represent the trained attribution model, while the grayish-lavender modules are auxiliary components for training. Data flows are indicated by different colors, with numbered arrows showing the order. 

The trained-model, built on MalConv-GCG \cite{ref3}, consists of: (1) a Word Embedding Layer for converting malware raw bytes into embeddings (\(x \rightarrow e\)); (2) a Representation Layer for extracting hidden vectors (\(e \rightarrow h\)); and (3) a Classification Head for predicting probabilities (\(h \rightarrow p\)). 
The auxiliary modules include: (1) a Projection Head for generating projection vectors (\(z\)); (2) an Adversarial Initialization module that creates initial adversarial samples ($x^{adv}$ in blue flow) using random perturbations; and (3) a Global Perturbation module that generates enhanced adversarial samples ($x^{adv}$ in green flow) by applying learned global perturbations.

The adversarial training of RoMA is formulated as the following min-max optimization problem, aiming to optimize the model parameters \(\theta\) and projection module \(\theta_P\) by minimizing the adversarial loss. Specifically, the inner maximization calculates the optimal perturbation \(\delta\) (using FGSM and GP) to maximize the adversarial loss on the perturbed sample \(x + \delta\), subject to the constraint \(\|\delta\| \leq \varepsilon\), where \(\varepsilon\) is the maximum perturbation magnitude. The outer minimization updates both \(\theta\) and \(\theta_P\) to minimize the total loss (incorporating adversarial consistency regularization to enhance robustness).
\begin{equation}
    \underset{\theta, \theta_P}{\min} \mathbb{E}_{(x, y) \sim D} \left[ \underset{\|\delta\| \leq \varepsilon}{\max} \ L_{CE} (\theta, x + \delta, y) \right]
    \label{equ_min_max}
\end{equation}


\begin{algorithm} 
\caption{Generating Adversarial Malware Samples: 
\(GenAdvMal\)\((x, y, \theta_{S}, W, \varepsilon, \eta, \mu)\).}\label{alg:alg1}
\textbf{Input}: clean malware sample and its group label \((x, y)\), select head’s parameter \(\theta_S\), word embedding matrix \(W\), maximum perturbation strength \(\varepsilon\), learning rate \(\eta\), momentum decay Coefficient \(\mu\).

\textbf{Output}: perturbed malware sample \(x^{adv}\).

\textbf{Global}: \(Pos\), perturbation-positions within malware \(x^{adv}\); $GP$, a set of global perturbation vectors, initialized as random embeddings; \(m\), A set of momenta for the gradients, initialized as zeros. 
\begin{algorithmic}[1]
\STATE \(x^{adv}\leftarrow Pack(x)\)
\STATE \(Pos\leftarrow\) Get all perturbation-positions in \(x^{adv}\)

\FOR {\(p\) in \(Pos\)}
\STATE \(x^{adv}[p]\leftarrow\) \textit{RandomByte()} 
\ENDFOR
\STATE \((e_1, h_1) \leftarrow model.Embedding.Representation(x^{adv})\)

\STATE \(z^\prime \leftarrow Selection(h_1)\)
\\ \hfill \textbf{\textit{\{FC head outputs $K$-dimensional logits $z^\prime$}}\textbf{\textit{\}}}
\STATE \(\theta_{S}\leftarrow \theta_{S}-\eta\nabla_{\theta_{S}}CLLoss(z^\prime, y)\) 
\vspace{0.05cm} %
\\ \hfill \textbf{\textit{\{Update Selection Head with CL loss\}}}
\STATE \(i \leftarrow \arg\max_{j \in \{1, \dots, K\}} z^\prime_j\) ~~~~~~~~~~~~~~~~~~~~~~~~~\textbf{\textit{\{Select $GP_i$\}}}
\vspace{0.05cm} %
\FOR {\(p\) in \(Pos\)}
\STATE \(e_1[p] \leftarrow e_1[p] + GP[i][p]\) 
~~~~~~~~~~~~~\textbf{\textit{\{Apply $GP_i$\}}}
\STATE \(x^{adv}[p]\leftarrow \operatorname*{argmin}_{j\in 0\dots255}\left ( \left\| e_{1}\left [ p \right ]-W_{j}\right\|_{2} \right )\)
\ENDFOR
\STATE \(e_2 \leftarrow model.Embedding(x^{adv})\) 
\STATE \(p_2\leftarrow model.Representation.Classification(e_2)\)
\STATE \(gradient\leftarrow \nabla_{e_2}CELoss(p_2, y) \)
\vspace{0.05cm} %
\FOR{\(p\) in \(Pos\)}
\STATE \(e_2[p]\leftarrow e_2[p] + \varepsilon * gradient[p]\)
\STATE \(x^{adv}[p]\leftarrow \operatorname*{argmin}_{j\in 0\dots255}\left ( \left\| e_2\left [ p \right ]-W_{j}\right\|_{2} \right )\)
\\ \hfill \textbf{\textit{\{Update malware content at perturbation-position}}\textbf{\textit{\}}}
\vspace{0.05cm} %
\STATE \(m[i][p] \leftarrow \mu * m[i][p] + \operatorname*{sign}(gradient[p])\)
\STATE \(GP[i][p] \leftarrow GP[i][p] + \varepsilon * \operatorname*{sign}(m[i][p])\)
\\ \hfill \textbf{\textit{\{Update $GP_i$\ in pool with momenta}}\textbf{\textit{\}}}

\ENDFOR

\RETURN \(x^{adv}\)
\end{algorithmic}
\end{algorithm}

\subsection{Adversarial Malware Generation with GP and FGSM}\label{subsec:adver}
The adversarial malware generation algorithm, \(GenAdvMal()\), is outlined in Algorithm \ref{alg:alg1}. RoMA first applies the UPX packing tool\footnote{https://upx.github.io/} to the clean malware \(x\) (line 1), producing a functionality-preserving compact version with an altered structure. This packed malware serves as the initial sample for adversarial perturbation. RoMA then identifies four types of \textit{Perturbation-Positions} and initializes them with random bytes (lines 2-5). \textbf{These positions} include parts of the DOS header excluding `MZ' and the PE pointer, a 1KB shifting space before the first section, unused slack space between sections, and up to 100KB of padding space at the file’s end, as \textbf{illustrated in Figure A1 (see Appendix)}. Next, RoMA refines the perturbation-position bytes using its key optimization technique (lines 6-22). 

RoMA's optimization approach is based on the assumption that \textbf{\textsl{malware adversarial samples with similar perturbations applied in the sample space may share a common perturbation pattern in the embedding space}}. Thus, as shown by the blue flow in Figure \ref{fig3}, RoMA first trains a Selection Head to adaptively choose a global perturbation (GP) for each malware in the embedding space, superimposes this GP onto the malware embedding and then adjusts the perturbation-position bytes in the sample space, generating an intermediate adversarial sample that aligns with its shared perturbation pattern (i.e., the GP). Subsequently, as indicated by the green flow in Figure \ref{fig3}, RoMA updates both the malware bytes at perturbation-positions and the GP based on FGSM, generating the final adversarial sample while refining the GP. The following subsections will detail these two steps.

\subsubsection{Selecting GP and Applying it at Perturbation-Positions}
RoMA utilizes a global perturbation (GP) pool consisting of $K$ GP vectors, each representing a shared perturbation pattern among malware embeddings. Each GP vector is initialized as a random embedding derived from random bytes at perturbation positions via the model’s word embedding layer, and is updated based on gradients (see next subsection). For each malware sample $x^{adv}$, after obtaining its embedding vector $e_1$ and representation vector $h_1$ through the model (line 6), RoMA trains a Selection Head to choose a GP from the pool. The Selection Head employs a single-layer fully-connected network, taking $h_1$ as input and producing a $K$-dimensional logits vector $z^\prime$ (line 7). The Selection Head is updated using the following contrastive learning objective (line 8):
\begin{equation}
    L^{\prime}_{CL}=\sum_{i=1}^{N}\frac{-1}{\left | P(i) \right | }\sum_{s\in P(i)}  log\frac{exp(z^{\prime}_i\cdot z^{\prime}_s/\tau)}{ {\textstyle \sum_{j\in N(i)} exp(z^{\prime}_{i}\cdot z^{\prime}_{j}/\tau )} } 
    \label{supcon}
\end{equation}
In each batch, for each anchor logits vector \(z^{\prime}_{i}\), $P(i)$ and $N(i)$ denote its sets of positive and negative vectors, respectively. Each positive vector \(z^{\prime}_{s}\) corresponds to a malware sample that shares the same APT-group label as the anchor malware (i.e., the malware corresponding to the anchor logits), while each negative vector \(z^{\prime}_{j}\)  corresponds to a malware sample with a different label. Thus, the CL loss optimizes the logits space by bringing the logits of in-group malware closer together and pushing those of out-group malware farther apart. 

The Selection Head selects the GP with the maximum value in the \textit{K}-dimensional logits vector (line 9). Intuitively, the selected GP represents the perturbation pattern with the highest logits score, indicating the most likely perturbation for the given malware. RoMA superimposes the selected GP onto the malware embedding $e_1[p]$, then adjusts the malware bytes at the \textit{perturbation-positions} in the sample space (by utilizing the closest word embedding vector in $W$), generating an intermediate adversarial sample $x^{adv}$ (lines 10-13).

\subsubsection{Updating Perturbation Bytes and GP via FGSM}

To maximize the likelihood of the adversarial malware being misclassified into an incorrect APT group, RoMA optimizes the perturbation bytes in the intermediate sample using FGSM, a gradient-based single-step attack strategy. Specifically, RoMA performs a forward pass of $x^{adv}$ through the model to obtain the embedding vector $e_2$ and the cross-entropy loss, then computes the gradient of this loss with respect to $e_2$ via backpropagation (lines 14-16). RoMA next calculates the embedding perturbation $\varepsilon * $\textit{gradient}$ $ based on the maximum perturbation strength and the gradient, and superimposes it onto the embedding vector $e_2$ (line 18). Finally, RoMA adjusts the perturbation bytes in the sample space to generate the final adversarial malware $x^{adv}$ (line 19).

To learn the GPs iteratively, RoMA synchronously updates the selected GP alongside applying gradient updates to each intermediate malware sample $x^{adv}$. Specifically, RoMA determines the perturbation direction for the GP using the sign of the gradient, and incorporates momentum to accumulate historical gradient information. (lines 20-21).


\subsection{Adversarial Training with Three Loss Functions}\label{subsec:loss}
RoMA employs fast adversarial training using single-step methods (e.g., FGSM) to defend against multi-step attacks. To enhance the effectiveness of single-step training, we propose two adversarial consistency regularization losses, $L_{AC}$ and $L_{AD}$, alongside the basic adversarial training loss, ($L_{AT}$), to optimize the representation quality and decision boundary. The Total loss is defined as below, where weight factors $\lambda _{1}$, $\lambda _{2}$ \(\in \left [0,1\right ]\):
\begin{equation}
        L_{Total}=L_{AT}+\lambda _{1}*L_{AC}+\lambda _{2}*L_{AD}.
    \label{equ6}
\end{equation}

As illustrated in Figure \ref{fig3}, the forward-pass data flows for each clean malware sample $x$ and its adversarial counterpart $x^{adv}$ are represented by black and red arrows, respectively. The model and the projection head produce the corresponding prediction vectors ($p$, $p^{adv}$) and projection vectors ($z$, $z^{adv}$), respectively. RoMA employs batch updates for both the model and the projection head, leveraging the cumulative effect of the three loss functions, which are detailed below:

\subsubsection{\textbf{Basic Adversarial Training loss (ATLoss)}}
\begin{equation}
    \begin{split}
        L_{AT} &= L_{AdvCE}+L_{CleanCE} \\
        &=-\frac{1}{N}\sum_{i=1}^{N}\sum_{j=1}^{G}\left(y_{ij}log\  p_{ij}^{adv}+y_{ij}log\ p_{ij}\right ).
    \end{split}
    \label{equ3}
\end{equation}
Here, $N$ is the batch size of clean malware, and $G$ denotes the number of APT groups. This loss combines the cross-entropy (CE) loss for both adversarial and clean malware samples, balancing the model's robustness and accuracy.

\subsubsection{\textbf{Adversarial Contrastive Loss (ACLoss)}} 
We propose an Adversarial Contrastive Loss to enhance malware representation by promoting intra-group perturbation invariance and inter-group distinguishability: \\
\begin{equation}
    \begin{gathered}
       L_{AC} = \frac{1}{2N}\sum_{i=1}^{2N}\frac{-1}{\left|CI(i)\right|+\left|AI(i)\right|}L_{s} , \qquad where  \\
         L_{s} = \sum_{s \in CI(i)\cup AI(i)}log\frac{e^{(z_{i}\cdot z_{s}/\tau)}}{e^{(z_{i}\cdot z_{s}/\tau)}+\sum_{j\in \textsl{N}_{i}}e^{(z_{i}\cdot z_{j}/\tau)}} \\
    \end{gathered}
\label{equ4}
\end{equation}
This loss incorporates both clean malware and their adversarial counterparts in the projection vector space. Each batch contains $2N$ samples: $N$ clean malware and $N$ adversarial variants. For each sample $x_i$, $z_i$ is its projection vector. $CI(i)$ is the set of Clean In-group instances with the same APT label as $x_i$, and $N_i$ is the set of Out-group instances, comprising all samples (both clean and adversarial) in the batch with a different label.
ACLoss treats the projection vectors of in-group samples (both clean and adversarial) as positive pairs, minimizing their distance. Conversely, out-group vectors are treated as negative pairs, maximizing their distance.

\subsubsection{\textbf{Adversarial Distribution Loss (ADLoss)}}
We propose an Adversarial Distribution Loss to minimize the divergence between the probability distributions of clean malware and its adversarial counterpart:
\begin{equation}
    \begin{split}
        L_{AD}=\frac{1}{N}\sum_{i=1}^{N}\sum_{j=1}^{G}\left (p_{ij}log\frac{p_{ij}}{p_{ij}^{adv}}\right ).
    \end{split}
    \label{equ5}
\end{equation}
This loss helps the attribution model establish a smoother and more consistent decision boundary, therefore enhancing its generalizability to both unseen and perturbed malware.

\section{Experiments}\label{sec5}
\subsection{Dataset and Evaluation Metric}

\quad \textbf{Dataset:} We introduce AMG18, a novel APT malware dataset comprising 6360 instances across 18 APT groups. We extend the APTMalware list (3594 samples from 12 groups \cite{ref57}) by incorporating additional hash-label pairs from public threat repositories. The group labels were verified by security analysts at a leading security firm in China, and the malware samples were obtained from the VirusSign database \cite{ref37}. \textbf{Table A1 in the Appendix outlines the details of AMG18}. Compared to APTMalware, AMG18 provides greater diversity with additional groups, realistic class imbalance, and over 100 samples per group, supporting robust splits and reducing overfitting risks. The dataset was split into an 80:20 train-test ratio for experiments.

\textbf{Evaluation Metrics:} We evaluate the effectiveness and robustness of each malware attribution model under both non-adversarial and adversarial settings. To reflect performance across all APT groups, we compute weighted metrics. Specifically, we use three metrics: (1) Standard Accuracy (SA), representing the accuracy on clean samples across all groups; (2) Robust Accuracy (RA), measuring the model's ability to correctly attribute adversarial samples; and (3) Attack Success Rate (ASR), indicating the rate at which attackers succeed in misleading the model (i.e., the effect of adversarial attacks). A higher SA indicates better performance in clean settings, while higher RA and lower ASR indicate greater robustness against adversarial attacks. \textbf{For detailed metric definitions, see Subsection C in the Appendix.}

\begin{figure*}
\centering
\subfloat[MalConv]{\includegraphics[width=0.24\textwidth]{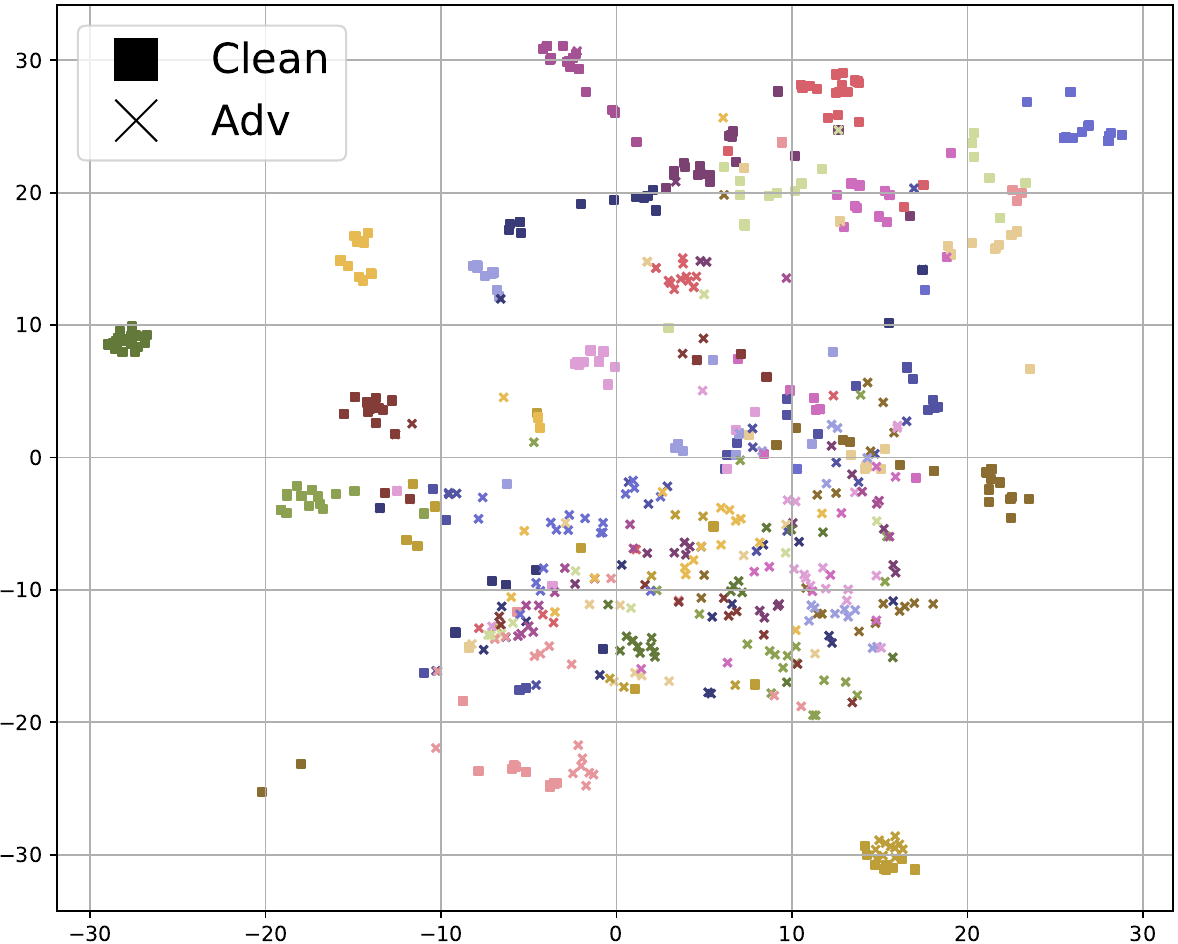}
\label{fig4_first_case}}%
\hfill
\subfloat[FGSM-AT]{\includegraphics[width=0.24\textwidth]{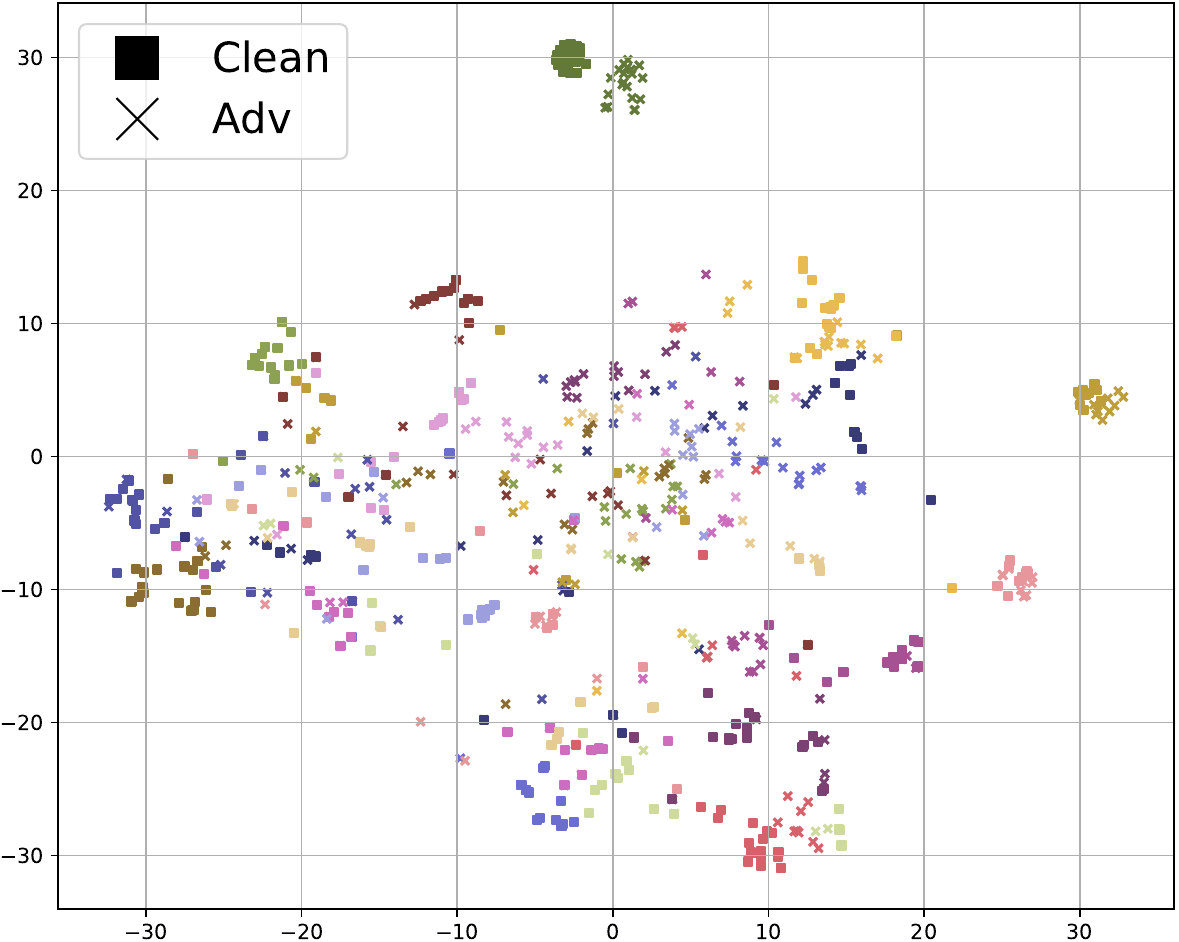}
\label{fig4_second_case}}%
\hfill
\subfloat[PGD-4-AT]{\includegraphics[width=0.24\textwidth]{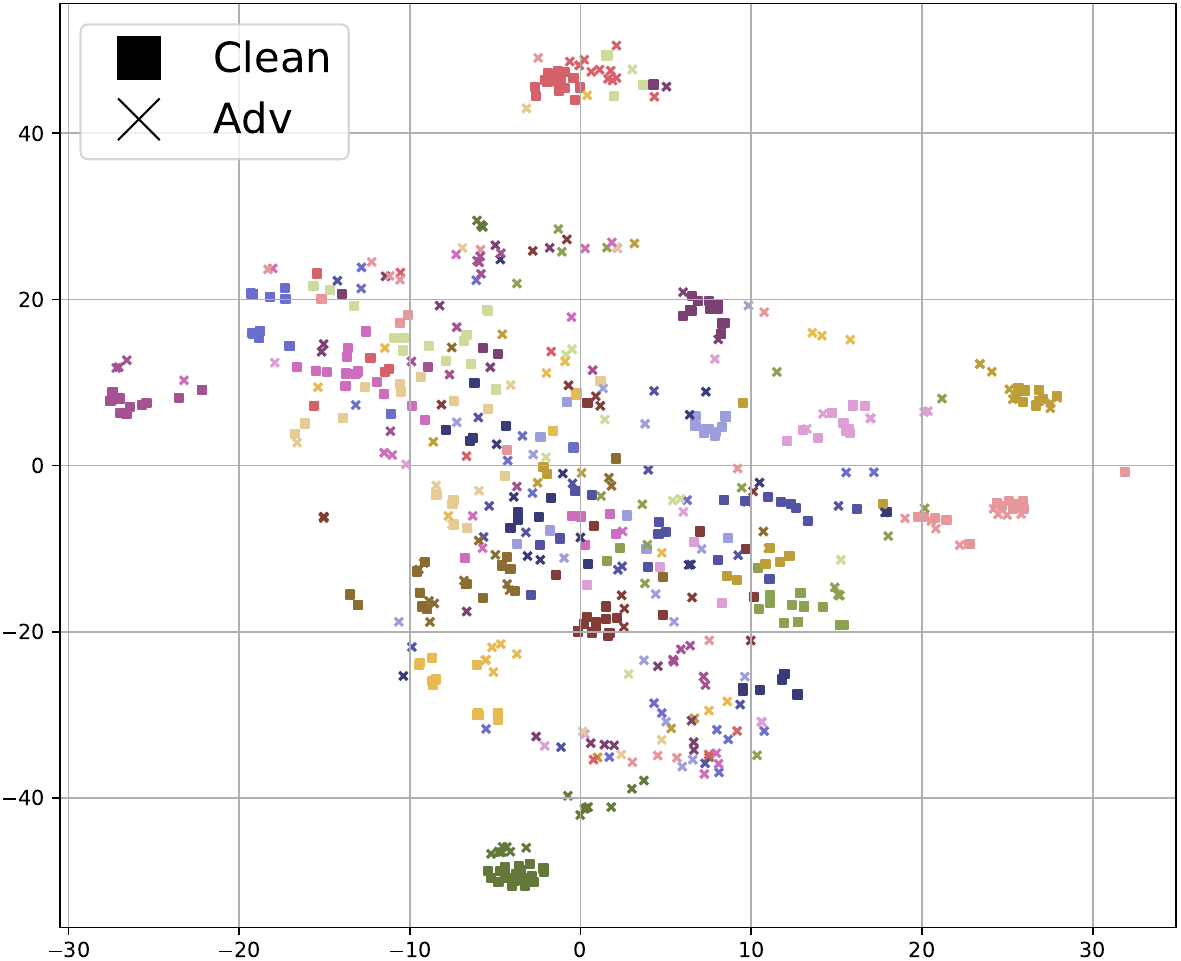}
\label{fig4_six_case}}%
\hfill
\subfloat[RoMA]{\includegraphics[width=0.24\textwidth]{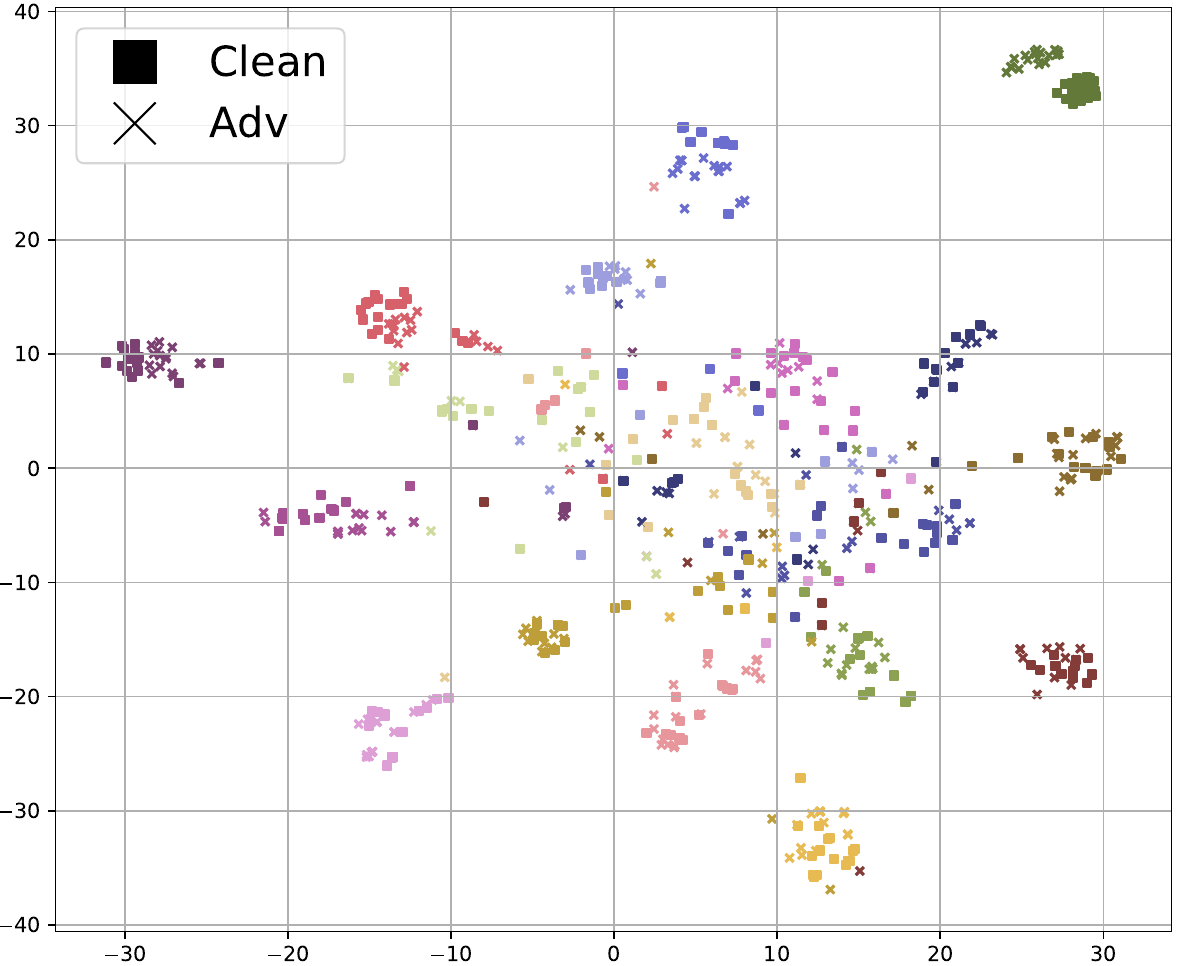}
\label{fig4_seven_case}}
\caption{Visualization of malware representation distributions for attribution models trained using four representative methods.}
\label{fig4}
\end{figure*}

\subsection{Experimental setting}\label{subsecB}
\subsubsection{\textbf{Attack Setting}}
For adversarial robustness evaluation, we employ two widely used attack methods: (1) PGD (Projected Gradient Descent), a multi-step iterative attack method \cite{ref55}, with a maximum perturbation strength \(\varepsilon = 0.6\) and 50, 60, or 70 iterations (PGD-50, PGD-60, PGD-70); (2) C\&W (Carlini \& Wagner), an optimization-based attack method \cite{ref50}.

\subsubsection{\textbf{Implementation Details}}
RoMA was implemented using PyTorch-Lightning 1.5.10, LIEF 0.10.0, and PyCharm as the IDE. The models were trained on a GPU server with dual RTX 4090 GPUs and Ubuntu 18.04.6 LTS. We used the Adam optimizer with a learning rate of 0.0001, and set the contrastive learning temperature \(\tau\) to 0.6. The loss weights $\lambda_{1}$ and $\lambda_{2}$ were both set to 0.3, while the number of GPs, \(K\), was set to 50, selected through extensive tuning. Training was conducted for 100 epochs with a batch size of 64.
     
\subsubsection{\textbf{Compared Methods}}
We compare the performance of the malware attribution models trained by RoMA against seven advanced competitor methods. Two non-adversarial methods: (1) MalConv \cite{ref3}: a SOTA PE malware classification method using the MalConv-GCG architecture for raw-byte analysis; (2) AvastNet \cite{ref58}: a neural network classification method for PE malware by analyzing raw bytes. One single-step adversarial training method: (3) FGSM-AT: an FGSM-based adversarial training method for PE malware classification, extending Slack-FGSM \cite{ref31} and Append-FGSM \cite{ref29} by incorporating RoMA's four perturbation positions (see Subsection \ref{subsec:adver}).
Two PGD-based multiple-step adversarial training methods: (4) PGD-2-AT and (5) PGD-4-AT \cite{ref56}: adversarial training implemented with two-step and four-step PGD attacks, respectively.
Two adversarial training methods adapted from image domains via transfer learning: (6) FGSM-RS \cite{ref49}: a single-step method enhancing FGSM with random initialization of adversarial perturbations; (7) NuAT \cite{ref42}: a single-step method incorporating a Nuclear-Norm regularizer.

\begin{table*}[] 
    \centering
    \setlength{\tabcolsep}{3.8pt}
    \begin{tabular}{lcrrrcccccc}
    \toprule
    \multirow{2}{*}[-1ex]{\textbf{Method}} & \multirow{2}{*}[-1ex]
    {\textbf{SA}}     & 
    \multicolumn{3}{c}{\textbf{RA}} & \multicolumn{3}{c}{\textbf{ASR}}  & \textbf{RA} & \textbf{ASR} & \multirow{2}{*}[-1ex]{\textbf{Time} \textit{(min)}} \\
    \cmidrule(r){3-5} \cmidrule(r){6-8} \cmidrule(r){9-9} \cmidrule(r){10-10} &  & \small\textit{\textbf{PGD-50}} & \small\textit{\textbf{PGD-60}} & \small\textit{\textbf{PGD-70}} & \small\textit{\textbf{PGD-50}} & \small\textit{\textbf{PGD-60}} & \small\textit{\textbf{PGD-70}} & \small\textit{\textbf{C\&W }}& \small\textit{\textbf{C\&W}} & \\
    \midrule
    MalConv & \underline{90.06} & 1.72 & 1.72 & 1.72 & 98.09 & 98.09 & 98.09 & 24.10 & 73.24 & 
    N/A 
    \\
    AvastNet & 89.67 & 1.41 & 1.41 & 1.41 & 98.43 & 98.43 & 98.43 & 23.32 & 74.00 & N/A \\ 
    FGSM-AT 
    & 87.87 & 10.33 & 9.86  & 9.62 & 88.25 & 88.78 & 89.05 & \cellcolor{gray!20}84.66 & \cellcolor{gray!20}3.65 & \underline{2706.57} \\
    
    PGD-2-AT \textit & 88.65 & 31.85 & 28.33 & 25.59 & 64.08 & 68.05 & 71.14 & \cellcolor{gray!20}86.70 & \cellcolor{gray!20}2.21
    & 3519.12 \\
    PGD-4-AT \textit & 89.59 & \underline{36.85} & \underline{34.90} & \underline{31.61} & \underline{58.86} & \underline{61.05} & \underline{64.72} & \cellcolor{gray!20}88.65 & \cellcolor{gray!20}1.05
    & 4965.48 \\
    FGSM-RS \textit{(TL)} & 89.44 & 11.50 & 9.15 & 6.89 & 87.14 & 89.76 & 92.30 & \cellcolor{gray!20}87.25 & \cellcolor{gray!20}2.45 
    & 2760.88 \\
    NuAT \textit{(TL)} & 89.51 & 1.80 & 1.80 & 1.80 & 97.99 & 97.99 & 97.99 & \cellcolor{gray!20}85.68 & \cellcolor{gray!20}4.28 & 2975.68 \\
   \midrule
    \textbf{RoMA }\textit{(Ours)} & \textbf{91.00} & \textbf{80.13} & \textbf{79.97} & \textbf{78.79} & \textbf{11.95} & \textbf{12.12} & \textbf{13.41} & \cellcolor{gray!20}88.02
    & \cellcolor{gray!20}3.27 & \textbf{2378.52} \\
	\bottomrule
    \end{tabular}
    \caption{Comparison of SA, RA, and ASR (\%) for Trained-Models across different Methods, along with adversarial training Time (minutes).}
    \label{tab1}
\end{table*}

\begin{table}[]
    \small
    \setlength{\tabcolsep}{3.4pt}
    \begin{tabular}{llll}
    \toprule
    \textbf{Method} & \textbf{SA ($\Delta$)}  & \textbf{RA ($\Delta$)} & \textbf{ASR ($\Delta$)} \\
     \midrule
    \textbf{w/o GP} & 90.68 ($\downarrow0.32$) & 14.16 ($\downarrow65.97$) & 84.38 ($\uparrow72.43$) \\
    \textbf{w/o ACLoss} & 84.42 ($\downarrow6.58$) & 55.71 ($\downarrow24.42$) & 33.98 ($\uparrow22.03$) \\
    \textbf{w/o ADLoss} & 89.98 ($\downarrow1.02$) & 79.03 ($\downarrow\enspace1.10$) & 12.17 ($\uparrow\enspace0.22$) \\

    \textbf{w/o AC\&AD} & 83.57 ($\downarrow7.43$) & 51.10 ($\downarrow29.03$) & 38.86 ($\uparrow26.91$) \\
    
    \textbf{w/o All} & 87.87 ($\downarrow3.13$) & 10.33 ($\downarrow69.80$) & 88.25 ($\uparrow76.30$) 
    \\
    \midrule
    RoMA \textit{(Ours)} & 91.00 ( / ) & 80.13 ( / ) & 11.95 ( / ) \\
    

	\bottomrule
    \end{tabular}
    \caption{Ablation study of RoMA method under PGD-50 Attacks.}
    \label{tab2}
\end{table}

\subsection{Clean Evaluation Without Attack}
Table \ref{tab1} presents the trained-model performance of RoMA and the compared methods. The second column, SA, reports the standard accuracy on the clean test set without attacks. Among all the trained-models, RoMA achieves the highest SA, followed by the SOTA MalConv. Notably, the non-adversarially-trained models of MalConv and AvastNet, outperform the adversarially-trained models in SA. This aligns with prior studies \cite{ref25,ref36}, which indicate that adversarial training, while enhancing robustness, often compromises performance on clean samples. However, RoMA stands out by not only improving robustness but also achieving superior performance under non-adversarial conditions.

\subsection{Robustness Evaluation Under Attacks}
Table \ref{tab1} presents the trained-model robustness of RoMA and seven competing methods against two attack types. (1) PGD Attacks: As the attack intensity increases from PGD-50 to PGD-70, performance deteriorates across all trained-models. However, RoMA consistently outperforms the others, maintaining a high RA of around 80\% and a low ASR of approximately 13\%. In contrast, the second-best PGD-4-AT achieves only around 36\% RA and over 58\% ASR. Notably, the two non-adversarially-trained models of MalConv and AvastNet experience a drastic drop, with RA below 2\% and ASR exceeding 98\%, making them extremely vulnerable to PGD attacks. 
(2) C\&W Attack: All adversarial-training methods exhibit strong resilience to the C\&W attack (RA $>$ 84\%, ASR $<$ 5\%) with minimal performance variation (highlighted in gray in the table). In contrast, the two non-adversarial-training methods experience a smaller decline compared to their performance under PGD attacks. This indicates that C\&W is not effective against adversarially-trained models, which is why this study focuses primarily on evaluating robustness under PGD attacks. In summary, compared to the competitors, RoMA demonstrates remarkably superior robustness against PGD attacks and strong resilience against C\&W attacks.

Moreover, the `Time' column in Table \ref{tab1} compares the adversarial training time of all the five adversarial-training methods. RoMA requires the least training time while achieving the best performance. In contrast, PGD-4-AT, which ranks second in performance, requires more than twice the training time.

\subsection{Ablation Study}
The proposed RoMA method incorporates the following key strategies to enhance the training of attribution models: Global Perturbation (GP) to generate stronger adversarial malware samples, and adversarial consistency regulation loss (comprising ACLoss and ADLoss) to improve malware representation. Table \ref{tab2} presents the ablation study results for removing these components, with $\Delta$ denoting the performance variation compared to models trained by the full RoMA.

Under clean conditions, as shown in the `SA' column, removing any single component causes a noticeable performance drop, with ACLoss contributing the most to improving clean accuracy. Under adversarial PGD-50 attacks, the `RA' and `ASR' columns reveal an even more pronounced decline. Specifically, removing GP dramatically reduces RA and substantially increases ASR, highlighting its critical role in training robust models; removing ACLoss significantly degrades robustness, while removing ADLoss causes a milder yet noticeable impact. The most severe degradation occurs when all three components are removed, with RA dropping by 69.80\% and ASR surging by 76.30\%. These results emphasize the importance of integrating all three components during training to achieve optimal model robustness.

\subsection{Visualization analysis}
To better analyze and visualize the performance of attribution models trained using RoMA and seven competitor methods, we applied t-SNE \cite{ref60} to project malware representation distributions into two dimensions. We selected 360 natural samples (20 per APT group) from the test set and generated corresponding adversarial samples, yielding 720 representation vectors. Figure \ref{fig4} shows the representation distributions for four representative methods: MalConv (non-adversarial), FGSM-AT (single-step), PGD-4-AT (multi-step), and RoMA. \textbf{Visualizations for all eight methods are included in Figure A2 in the Appendix.} In these figures, different APT groups are distinguished by unique colors, with clean samples shown as solid-squares and adversarial samples as cross-signs.

As the figures show, the model trained with RoMA demonstrates superior representation distribution compared to all competitors, with closer within-class clustering  (particularly between adversarial and clean samples) and improved between-class separation. The clearer decision boundary between classes further highlights its effectiveness. RoMA achieves this by generating stronger adversarial malware through GP, enhancing within-class clustering and between-class separation via ACLoss, and refining the decision boundary with ADLoss.

\section{Conclusion}\label{sec6}
This work addresses the challenge of robust APT malware group attribution through byte-level analysis. We proposed RoMA, a fast adversarial training approach that integrates global perturbations and adversarial consistency regularization, achieving significant improvements in robustness. To support evaluation, we introduced AMG18, a novel APT malware dataset with group labels. Experimental results demonstrate that RoMA consistently outperforms seven competitor methods with remarkable robustness, high accuracy, and efficient training. Future work includes applying RoMA to diverse malware types, additional classification tasks, and broader domains. The RoMA-trained model and visualized malware samples are publicly available. 

\bibliographystyle{named}
\bibliography{ijcai25}

\newpage
\appendix

\renewcommand{\thefigure}{A\arabic{figure}}
\renewcommand{\thetable}{A\arabic{table}}

\section{Perturbation Positions in PE File}
\begin{figure} [H]
\centering
\includegraphics[width=0.48\textwidth]{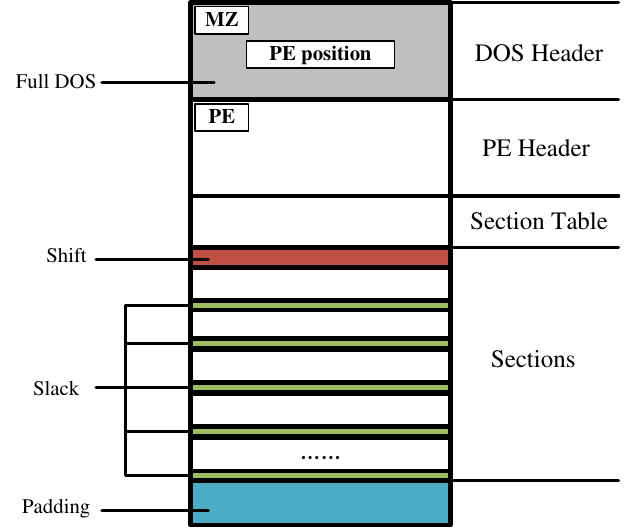}
\centering
\caption{Perturbation positions in a PE file}
\label{fig2}
\end{figure}

Figure \ref{fig2} illustrates four types of perturbation positions (Perturb-Pos) within a PE file: (1) Full DOS Region: All positions in the DOS header, excluding the 'MZ' magic number and the 4-byte pointer to the PE header; (2) Shifting Space: A 1KB gap created by shifting the first section relative to the PE header; (3) Slack Space: Unused regions between sections; (4) Padding Space: Extra space appended at the end of the PE file, up to 100KB.

\section{Dataset}   
    Table \ref{tab1} details the 18 APT groups in our dataset and their respective malware sample counts. This dataset exhibits class imbalance, reflecting real-world scenarios. Every class in our dataset contains over 100 samples, ensuring adequate data for a robust train/test split and reducing the risk of overfitting due to limited training samples.

\begin{table}[h]
    	\centering
    	\caption{APT Malware Dataset}
    	\begin{tabular}{cccc}
    	\toprule
    	\textbf{GroupName} & \textbf{Count} & \textbf{GroupName} & \textbf{Count}\\
    	\midrule
    	APT17              & 1428           & Winnti             & 245            \\
    	Lazarus Group      & 888            & APT29              & 211            \\
    	OceanLotus         & 512            & Blackgear          & 186            \\
    	APT10              & 507            & Donot              & 180            \\
    	Equation Group     & 395            & TA505              & 147            \\
    	APT1               & 345            & BITTER             & 135            \\
    	PatchWork          & 315            & Energetic Bear     & 124            \\
    	BlackTech          & 280            & APT28              & 109            \\
    	Darkhotel          & 251            & APT30              & 102            \\
    	\bottomrule
    	\end{tabular}
    	\label{tab1}
\end{table}

\section{\textbf{Evaluation Metrics}}
\begin{enumerate}
        \item[a)] \textbf{Standard Accuracy (SA)}: SA measures the model's attribution accuracy on clean samples across all groups. 
        \begin{equation}
            SA=\sum_{i=1}^{n}\left ( \frac{C_{i}^{clean}}{T_{i}^{clean}} \right )\times \left ( \frac{T_{i}^{clean}}{\sum_{j=1}^{n}T_{j}^{clean}}\right ),
            \label{equ7}
        \end{equation}
where \(n\) is the number of attributed APT groups, \(C_{i}^{clean}\) is the number of correct attributions for group \(i\) on clean samples, \(T_{i}^{clean}\) represents the total number of clean samples for group \(i\), and \(\frac{T_{i}^{clean}}{\sum_{j=1}^{n}T_{j}^{clean}}\) denotes the weight for group \(i\) among the clean-sample groups.
        \item[b)] \textbf{Robust Accuracy (RA)}: RA evaluates the model's robustness in correctly attributing adversarially perturbed samples across all groups. 
        \begin{equation}
            RA=\sum_{i=1}^{n}\left ( \frac{C_{i}^{adv}}{T_{i}^{adv}} \right )\times \left ( \frac{T_{i}^{adv}}{\sum_{j=1}^{n}T_{j}^{adv}}\right ),
            \label{equ8}
        \end{equation}
where \(C_{i}^{adv}\) is the number of correct predictions for group \(i\) on adversarial samples, \(T_{i}^{adv}\) is the total number of adversarial samples for group \(i\), and \(\frac{T_{i}^{adv}}{\sum_{j=1}^{n}T_{j}^{adv}}\) represents the weight for group \(i\) in the adversarial-sample groups. 
     \item[c)] \textbf{Attack Success Rate(ASR)}: ASR is one of the most direct indicators to measure the effect of adversarial attacks. It represents the rate at which the attacker succeeds in spoofing the model and obtaining the wrong output. A higher ASR means that adversarial attacks are more effective and the model is less robust against such attacks.
     \begin{equation}
         ASR=\sum_{i=1}^{n}\left ( \frac{C_{i}^{clean}-C_{i}^{adv}}{C_{i}^{clean}} \right )\times \left ( \frac{C_{i}^{clean}}{\sum_{j=1}^{n}C_{j}^{clean}}\right ),
         \label{equ9}
     \end{equation}
     where \(\frac{C_{i}^{clean}}{\sum_{j=1}^{n}C_{j}^{clean}}\) denote the weight for correct predictions of group \(i\) in the correct predictions of clean-sample groups.
\end{enumerate}

\section{\textbf{Visualization Results}}
\renewcommand{\thefigure}{A\arabic{figure}}
The representation visualization is shown in Figure \ref{fig4}.

\begin{figure*}
\centering
\subfloat[MalConv]{\includegraphics[width=0.41\textwidth]{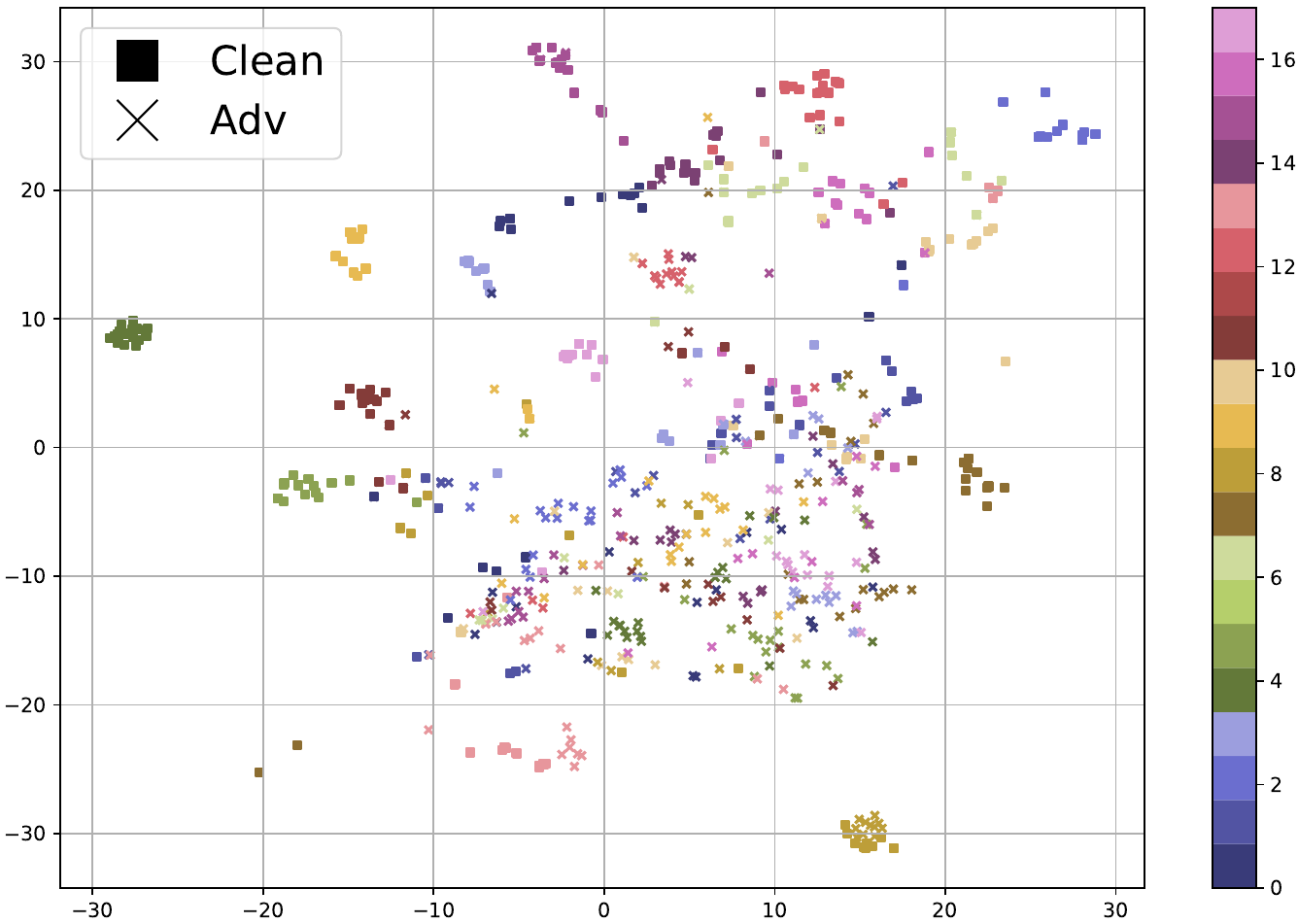}
\label{fig4_first_case}}%
\hfill
\subfloat[AvastNet]{\includegraphics[width=0.41\textwidth]{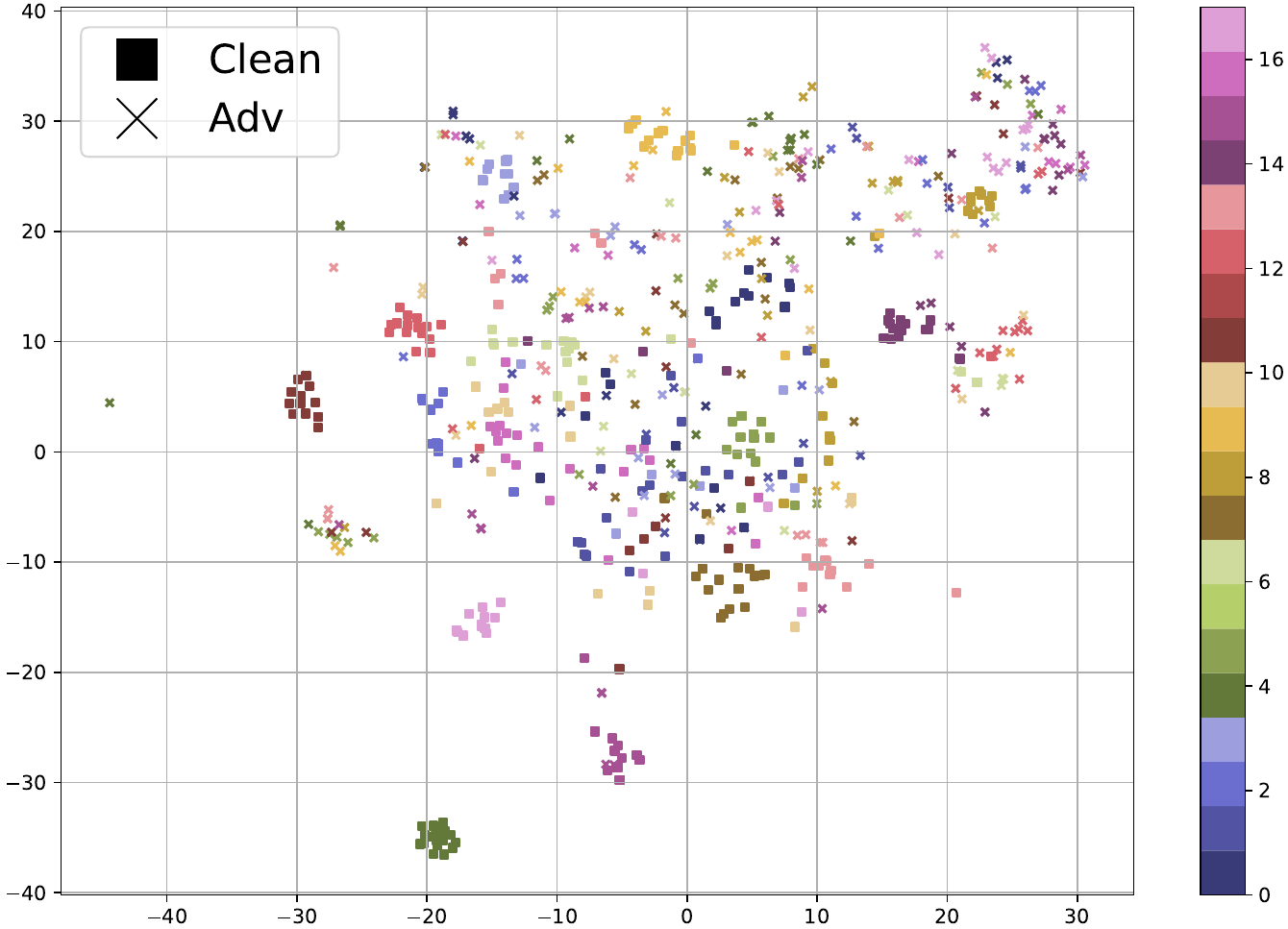}
\label{fig4_eight_case}}%
\hfill
\subfloat[FGSM-AT]{\includegraphics[width=0.41\textwidth]{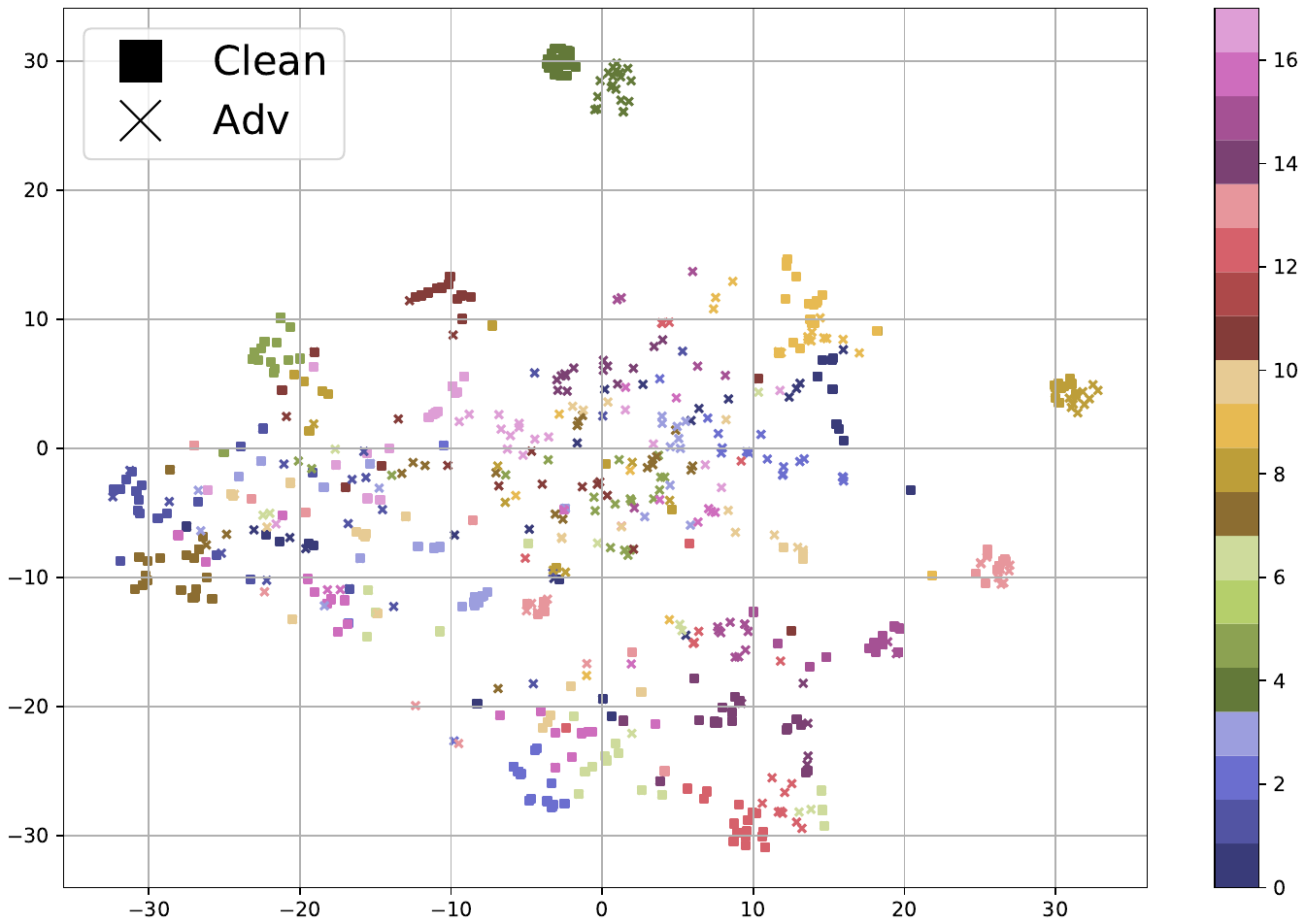}
\label{fig4_second_case}}%
\hfill
\subfloat[FGSM-RS]{\includegraphics[width=0.41\textwidth]{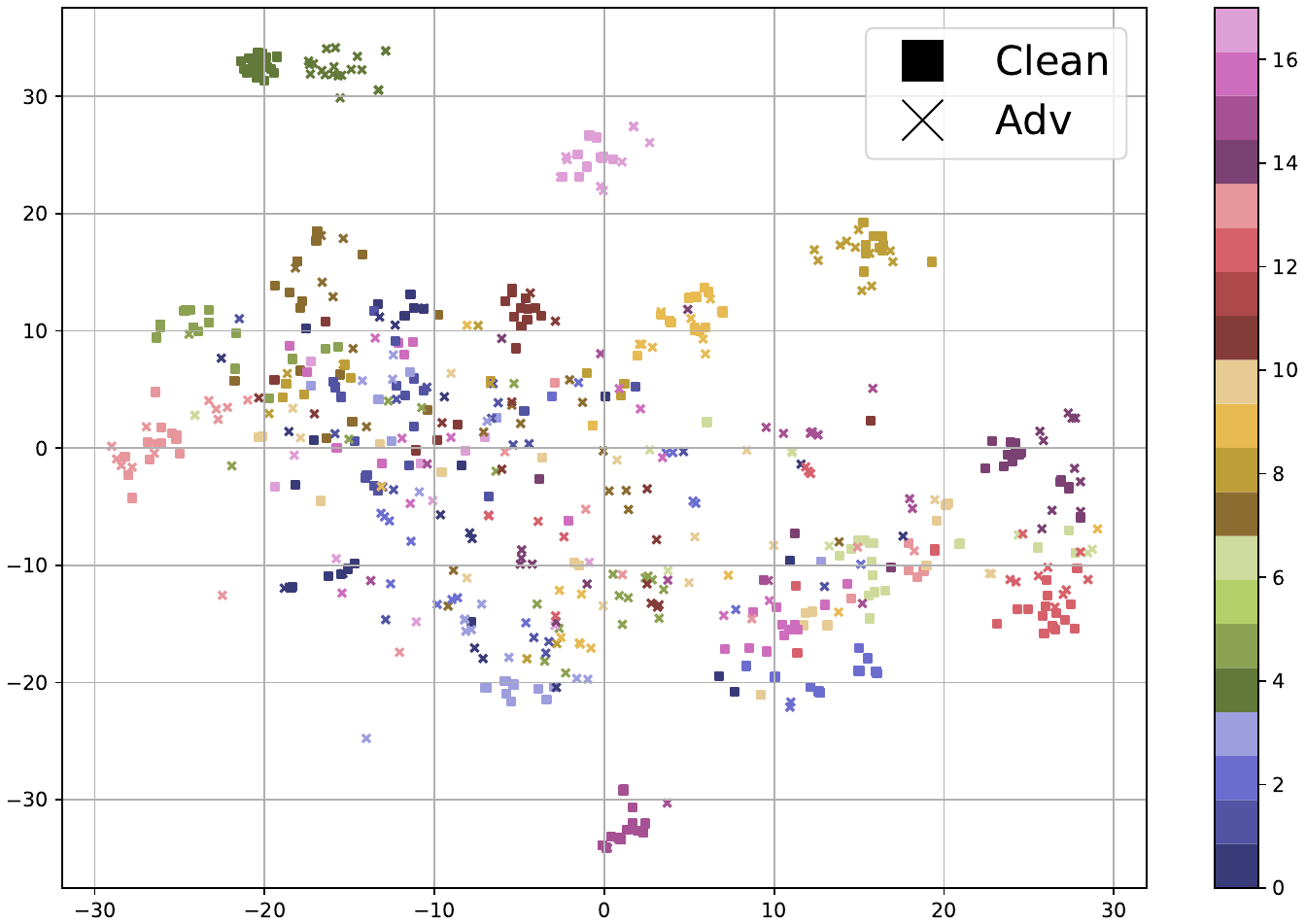}
\label{fig4_third_case}}%
\hfill
\subfloat[NuAT]{\includegraphics[width=0.41\textwidth]{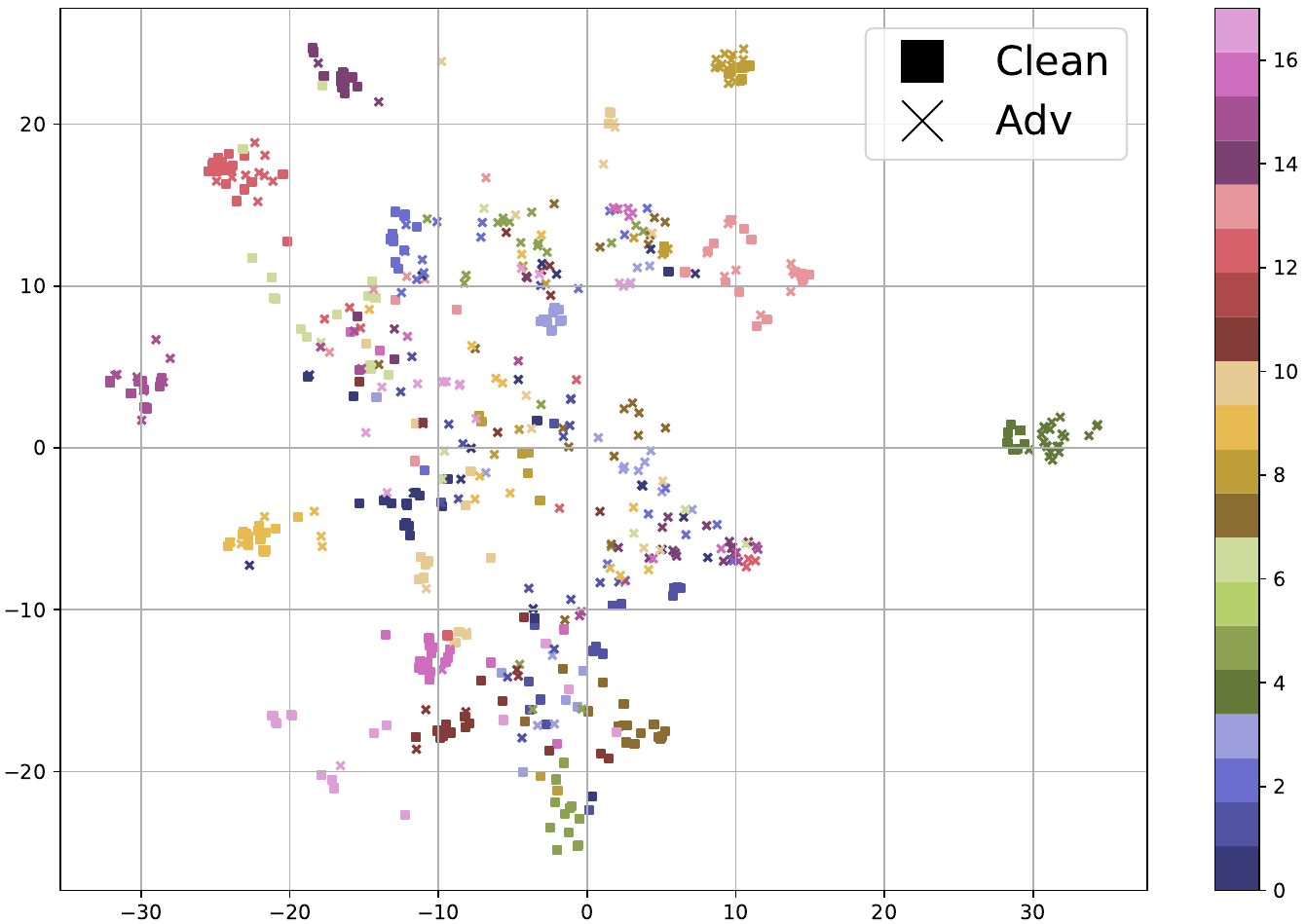}
\label{fig4_five_case}}%
\hfill
\subfloat[PGD-2-AT]{\includegraphics[width=0.41\textwidth]{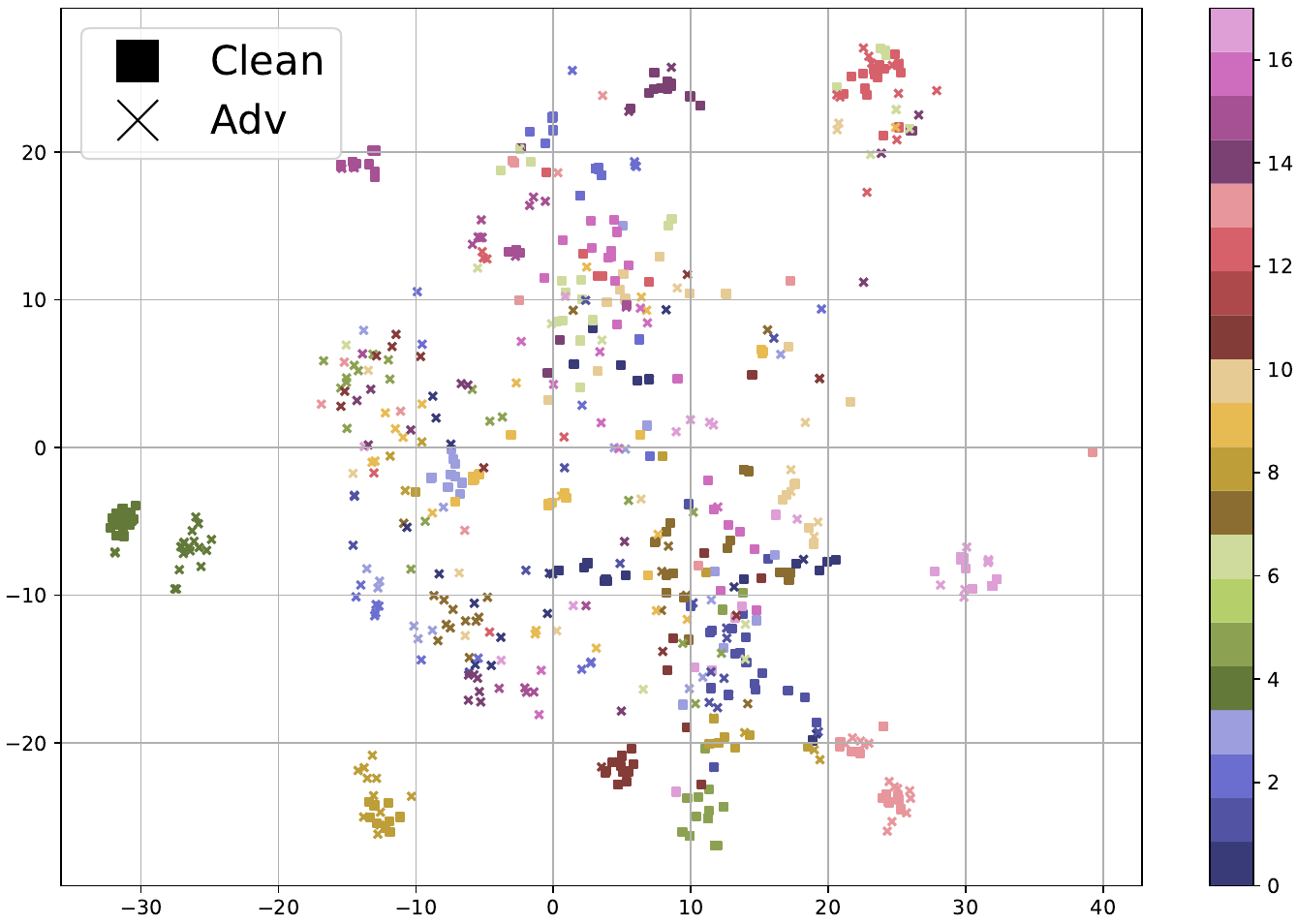}
\label{fig4_six_case}}%
\hfill
\subfloat[PGD-4-AT]{\includegraphics[width=0.41\textwidth]{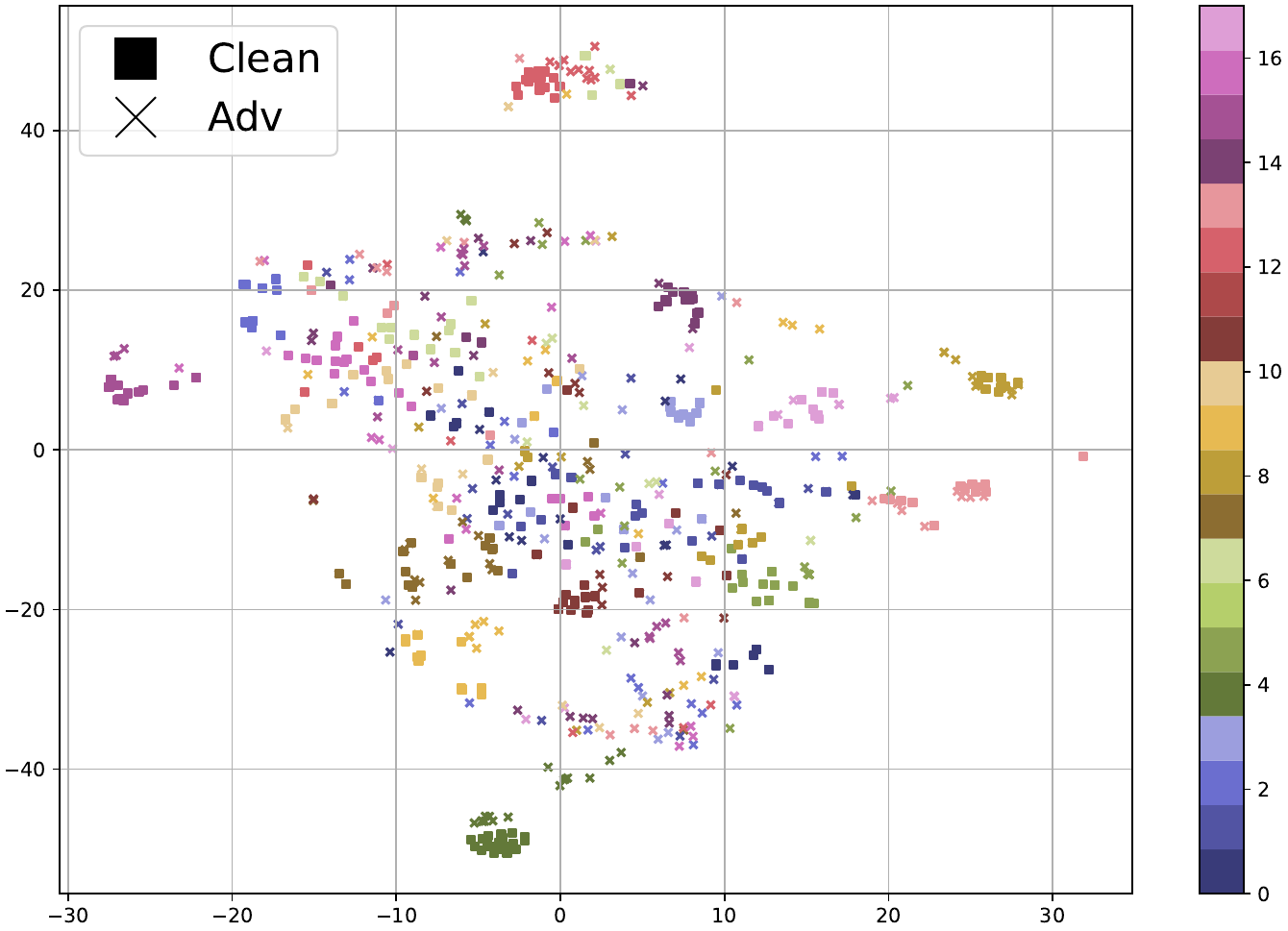}
\label{fig4_six_case}}%
\hfill
\subfloat[RoMA]{\includegraphics[width=0.41\textwidth]{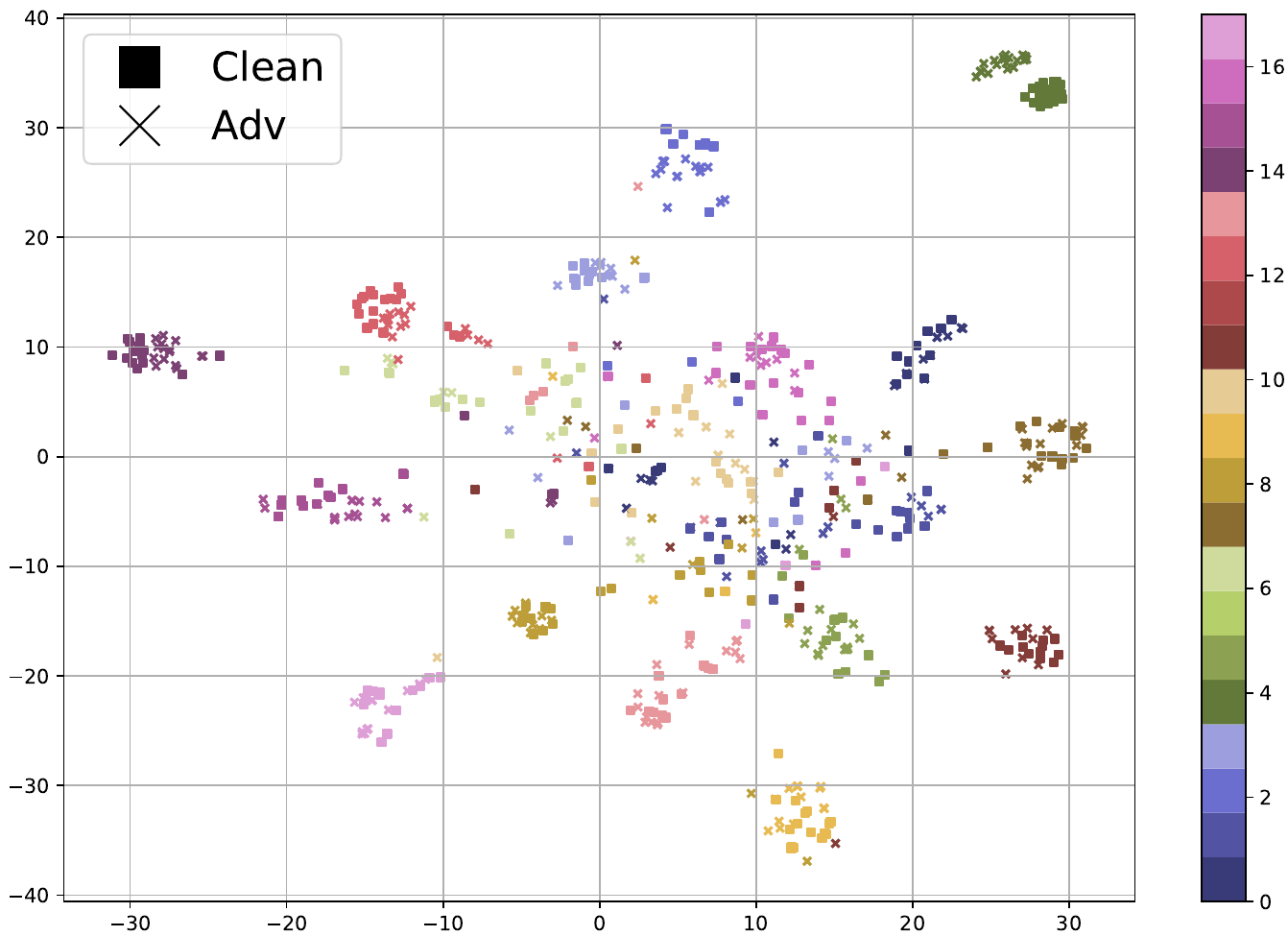}
\label{fig4_seven_case}}
\caption{Visualization of malware representation distributions for attribution models trained using all eight methods.}
\label{fig4}
\end{figure*}

\end{document}